# A Tunable Electrorheological Fluid Microfluidic Rectifier: Irreversibility of Viscous Flow due to Spatial Asymmetry Induced Memory Effects

X. Huo[1] and G. Yossifon[1*]

[1]*Faculty of Mechanical Engineering, Micro- and Nanofluidics Laboratory, Technion–Israel Institute of Technology, Technion City 32000, Israel*

* Corresponding author: yossifon@technion.ac.il

Due to the reversibility of viscous flow it is not expected to obtain a fluidic rectifier simply from geometrical asymmetry without any moving mechanical parts. Here, we found a counter example by using spatial asymmetry combined with electric field to inject memory effects that render the flow irreversible. This stems from the strong dependency of the electrorheological (ER) fluid particle chaining on the flow direction. A funnel-shaped microfluidic rectifier with ER fluid has been shown to be easily and rapidly tuned via the applied electric field to achieve more than an order of magnitude rectification along with pressure oscillations. These findings are of importance for realization of fluidic diodes, rectifiers and ratchets.

Electrorheological (ER) fluids consist of solid particles dispersed in an insulating carrier fluid, mostly dielectric oils or organic liquids. Under applied electric fields, the particles become electrically polarized and align themselves to form chains and columns along the field direction, which cause a dramatic increase in apparent viscosity of the ER fluid, resulting in a phase transition from a liquid to a solid-like state. The phase transformation is achieved very quickly, usually on the order of 1-10 ms, and is reversible. This unique feature renders ER fluids promising agents in electric-mechanical interfaces such as shock absorbers, breaks, clutches, valves and actuators [1].

Microfluidics has emerged and prospered for over one decade as a low-cost technique for precise spatial and temporal control of analytes and rapid biological or chemical diagnosis [2]. It is applied in an enormous number of applications in point-of-care (POC) and lab-on-a-chip (LOC) systems. ER fluids have gained increasing interest in microfluidics, due to their diverse microfluidic applications, e.g., microvalves [3,4], micropumps for driving flow in a programmable manner [5] and even logic control of microfluidics [6].

Inspired by the effects of symmetry-breaking geometries of ionic-permselective pores (e.g. funnel-shaped nanochannels [7] or conical-shaped pores [8]) on ionic current rectification, reminiscent of electron rectification in diodes, we set out to examine the effect of similar symmetry breaking on ER fluid behavior. However, due to the reversibility of viscous flow at low Reynolds numbers [9] it is not expected to obtain a fluidic diode, rectifier or ratchet [10], wherein the resistance depends on the direction of flow, simply from geometrical asymmetry of the channel. Here we found a counter example by using spatial asymmetry to inject memory effects, stemming from flow direction dependency of electrorheological fluid particle chaining, that render flow irreversible. Electrorheological fluid behaves Newtonian in the absence of an electric field and Bingham in the presence of an electric field [11-13].

Flow irreversibility in the low Reynolds regime was previously reported for non-Newtonian viscoelastic fluids with triangular [14, 15] and hyperbolic [15] nozzle/diffuser shapes as well as contraction with an obstacle design [16] in order to achieve high anisotropic flow resistance between the two flow directions. However, the fluidic diodicity or rectification factor is quite low, in the range of 1.5-6, and is an inherent fixed parameter of the system for a given flow rate. Previously reported microfluidic rectifiers with Newtonian fluid involved complex channel design that included moving mechanical parts, e.g. valve [17] or flap [18].

To the best of our knowledge the current study is the first realization of a fluidic rectifier without moving mechanical parts which can be dynamically and rapidly tuned via the applied electric field to achieve more than an order of magnitude rectification. Furthermore , all previous studies on ER fluids examined straight and parallel actuating electrodes. Here, we broke the parallel electrode setup commonly used in ER studies via a funnel-shaped microchannel obtained by simply tilting the electrodes side walls (Fig.1a, b). It is obvious that in such a geometry, where the gap between the opposing electrodes varies from tip to base, and in turn, the electric field intensity, the strength and density of the ER chains also vary along the channel. In such a symmetry-broken channel, it is expected that the ER effect will depend upon the flow direction, hence, leading to a rectification effect, manifested via effective viscosity instead of current.

Fabrication of the chip followed the technique used by Helal et al. [19], wherein the channel geometry was obtained using direct laser micromachining (Mob laser 50, with L Designer) of a copper foil (3M, 1182, 90 μm in thickness) and Kapton tapes arranged on a glass slide, where the middle section of copper foil served as the electrodes (Fig.1a). The channel was sealed with another glass slide, with an intermediate layer of 50μm thick adhesive tape (3M, 467MP). The copper foils were connected to a function generator (TTi, TGA12104) in series with a high-voltage amplifier (Trek, 2220). The ER fluid examined was RheOil4, with an average diameter of 3.5μm polyurethane particles in silicone oil at a volumetric concentration of $\phi=3\%$ Vol, which was obtained upon dilution with 50 cSt silicone oil (Sigma-Aldrich). The funnel-shaped microfluidic chip was mounted on an inverted epi-fluorescent microscope (Nikon, Eclipse Ti-U) for visualization. The ER fluid was then loaded into a syringe and introduced into the microchannel via a tubing (Qosina, Tygon), with flow rate, $Q$, controlled by a syringe pump (KDS Legato 210). The pressure drop across the microchannel inlets was measured (using Honeywell SSCDANT030PG2A3 pressure sensors) for varying volumetric flows and electric field intensities.

In the current study, we chose to apply an electric field of a square-wave form at a frequency of 10Hz, where electrophoresis effects become negligible, resulting in the symmetric accumulation of the particles at the two opposite electrodes (See Fig. S1). In contrast, under DC conditions, the particles, having a native charge, accumulated on one of the electrodes, thus, leading to non-symmetric flow and to degradation of the microfluidic device due to non-reversible absorption of the particles at the electrodes edge. Interestingly, we obtained a maximum in the ER response (Fig.S1) at ~100Hz, followed by high frequency dispersion. Previous studies either reported an almost constant apparent viscosity up to ~100Hz [20] or also noted the existence of such a maximum [21]. Although interesting, the detailed study of the effect of frequency on the ER response is left for future studies and here, we fixed the frequency to 10Hz. Owing to the design of the microchannel device, high-quality visualization of the temporal behavior of the particles was feasible and was synchronized with the pressure sensors and function generator signals.

The flow direction from base to tip was defined as positive, while from tip to base was defined as negative. For a positive flow direction, particle feeding from the funnel base to tip further intensified particle blockage at the funnel tip, leading to a rapid buildup of pressure (Fig.1c) and vice-versa for the negative flow direction, wherein tip blockage was delayed to higher voltages due to the counteracting shearing effect of the flow (Fig.1d). This was also manifested by a transient response of the pressure drop for the positive (Fig.2a) and negative (Fig.2b) flow directions. At sufficiently low voltages, where there was no particle blockage of the tip, the pressure drop saturated at a low value. Interestingly, we noted that oscillations of the pressure response occurred beyond the critical voltage for particle blockage of the funnel tip, $V_c$, which was associated with a rapid buildup of pressure and particle accumulation. At a certain peak pressure, defined as the yield pressure [19], the particle blockage destabilized and broke, as indicated by the sharp pressure drop which was again restored and broke in a periodic manner.

Plotting the maximum pressure drop response (average of the maximum pressure values in an oscillatory response, as indicated in Fig. 2b) for both directions (Fig. 2c) clearly showed delayed critical voltage for the particle blockage for the negative flow direction. This dependency on the flow direction can be described via a hydrodynamic rectification factor defined as the ratio between the pressure drops in the positive versus negative direction for the same applied voltage and flow rate. The difference between the critical voltage for the positive and negative flow directions (marked respectively as $V_{c,p}$ and $V_{c,n}$ in Fig.2c), resulted in a more than one order of magnitude ER rectification (Fig.2d), with a distinct maximum corresponding to the delayed critical voltage of the negative direction. The maximum was due to the fact that at voltages sufficiently below the critical voltage, the differences between the positive and negative flow directions becomes negligible and hence, the rectification factor becomes unity. The same applies to sufficiently large voltages, where the ER effect overwhelms the hydrodynamic shearing effect and becomes independent of the flow direction, hence, the maximum appears only at intermediate voltages at which the these two effects are comparable (see Fig.S2 and S3).

For simple scaling analysis, we used the dimensionless Mason number [22], $Mn$, that characterizes the relation between hydrodynamic shear and electric polarization forces

$$Mn = \frac{\eta_c \dot{\gamma}}{2\varepsilon_0 \varepsilon_c (\beta E)^2}, \qquad (1)$$

where $\eta_c$ is the viscosity of the continuous phase, $\dot{\gamma}$ is the shear rate, $\varepsilon_o$ and $\varepsilon_c$ are the permittivity of the vacuum and the dielectric constant of the continuous phase, respectively, $E$ is the electric field strength, and $\beta$ is the relative polarizability of the particles [22].

The critical electric field beyond which the tip region is blocked, is taken as $E_c = V_c/h$. Since at voltages around the critical voltage the hydrodynamic and electric polarization forces acting on the chains are comparable, we can use the criteria of $Mn \sim 1$ to approximate the critical electric field. In a channel with a rectangular cross-section, the shear rate can be described as [23]

$$\dot{\gamma} = \frac{6Q}{hw^2}\left(1 + \frac{w}{h}\right) f, \qquad (2)$$

where $f$ is a decreasing function of $w/h$. Combining Eqs. (1) and (2) and $Mn \sim 1$ yields

$$V_c = K\left(Q(h+w)f\right)^{0.5}, \qquad (3)$$

where $K$ is the scale factor. For a fixed $Q$, this scaling seems to qualitatively predict the experimental results (Fig. 2d as well as Fig.S4, depicting similar behavior, for a microchannel with a tip opening of 400µm) in terms of the dependency of the voltage at which the rectification maximum occurs, $V_{max}$ ($\approx V_{c,n}$), on the tip opening, $h$, while the fitted prefactor $K = 3.9 \cdot 10^9 Vm^{-2}s^{0.5}$.

We then further examined the effect of the flow rate on ER rectification by measuring the pressure drop versus voltage for both positive and negative flow directions with varying flow rates (Fig.3a). As clearly seen, the critical voltages for funnel tip blockage for both the positive and negative directions, shifted to higher voltages with increasing flow rates (Fig.3a). The voltage at which the rectification factor maximum ($V_{max} \approx V_{c,n}$) occurred exhibited a similar trend (Fig.3b). From Eq.(3), for fixed channel dimensions, $h$ and $w$, the critical voltage ($V_{max} \approx V_{c,n}$) scales with $Q^{0.5}$ is indicated by the dashed line in the inset of Fig.3b, with a fitted prefactor $K = 4.1 \times 10^9 Vm^{-2}s^{0.5}$.

The time period, $T$, of the pressure oscillations (indicated in Fig. 2b) is expected to linearly scale with the electric field squared, regardless of the direction of the flow, in accordance with the scaling of the particle-particle force as seen in Fig. 3c. Since the pressure oscillations correspond to the periodic buildup and destruction of the accumulated particle chains within the funnel, the funnel can be treated as a container that destabilizes once reaching its finite filling capacity. Accordingly, the filling time should scale inversely with the flow rate, $Q$. Expressing these relations with respect to the threshold values of the electric field, $E_{th}$, and its associated periodic time, $T_{th}$, for the occurrence of the oscillations yields the following universal scaling

$$T - T_{th} \propto \left(E^2 - E_{th}^2\right)/Q, \qquad (4)$$

as depicted in the inset of Fig. 3c. It is seen that the periodic time for the positive flow is higher than the negative flow direction in accordance with the higher yield pressure for the former.

The mechanism underlying the rectification effect is described in Fig.4. The asymmetry in the ER effect cannot simply arise from the hydrodynamics, since the solution to the Stokes flow (Reynolds number $\mathrm{Re} = \rho u h/\eta \sim 0.05$; where the density is $\rho = 1.03$ kg/l, the dynamic viscosity is $\eta = 0.035$ Pa·s, the width of the opening is $h$=150μm, and the velocity is $u = 0.012$ m/s for $Q$=10μL/min) is symmetric upon change of the flow direction. The same applies for the electric field, which is the solution of the Laplace equation. However, the duration of time the particles have to form chains in the vicinity of the funnel tip, performing as a positive DEP (p-DEP) trap, is very different for the opposite flow directions. When the flow is in the positive direction, the chains have sufficient time to form along the funnel length. However, when the flow is in the negative direction, the particles begin to interact with each other, and begin to form chains already at the inlet to the tip of the funnel. If the distance at which the chains are formed, $s$ (Fig.4a), is larger than a critical value, $s_c$, then the chains escape the p-DEP trap (Fig.4b) and there is no jamming of the funnel, which, in turn, results in an intensified ER effect. It is expected that the distance at which chain formation initiates, depends on the inverse of the strength of the particle-particle force, i.e. $s \propto E^{-2}$ (Fig.4c).

To summarize, a significant rectification, of more than one order of magnitude, of the apparent viscosity was obtained due to an ER effect upon symmetry-breaking of the electric field using non-parallel electrodes with a funnel-shaped microchannel. A distinct maximum rectification was observed, with a corresponding voltage (approximately the critical voltage for the blockage of the tip region in the negative flow direction) that monotonically increased with increasing $h$ and/or $Q$, according to the simple scaling expression of Eq.(3). It was demonstrated that spatial asymmetry injected memory effects, via strong flow direction dependency of the particle chaining, that render flow irreversible.

Such a diode-like ER device is of both fundamental and applicative importance and may stimulate further studies of the effect of symmetry breaking as well as introduce new functionalities to ER-based microfluidic and micro-electro-mechanical devices. For example, symmetry breaking geometries could be useful for realization of ER fluidic equivalents of electrical circuit elements. The latter has been the focus of intensive research efforts in recent years using nanofluidics, e.g., diode [8], transistors [24] and recently also using ER-based droplet microfluidics [6,25]. This may ultimately lead to large-scale integration of microfluidic processors with more complexity for on-chip processing for chemical and biological reactions and testing, drug screening, and others.

We wish to acknowledge MAFAT for their financial support and the Technion Russel-Berrie Nanotechnology Institute (RBNI) and the Technion Micro-Nano Fabrication Unit (MNFU) for their technical support. We also thank Prof. Touvia Miloh, Prof. Steve Frankel, Dr. Nipun Arora and Dr. Sinwook Park for helpful discussions.

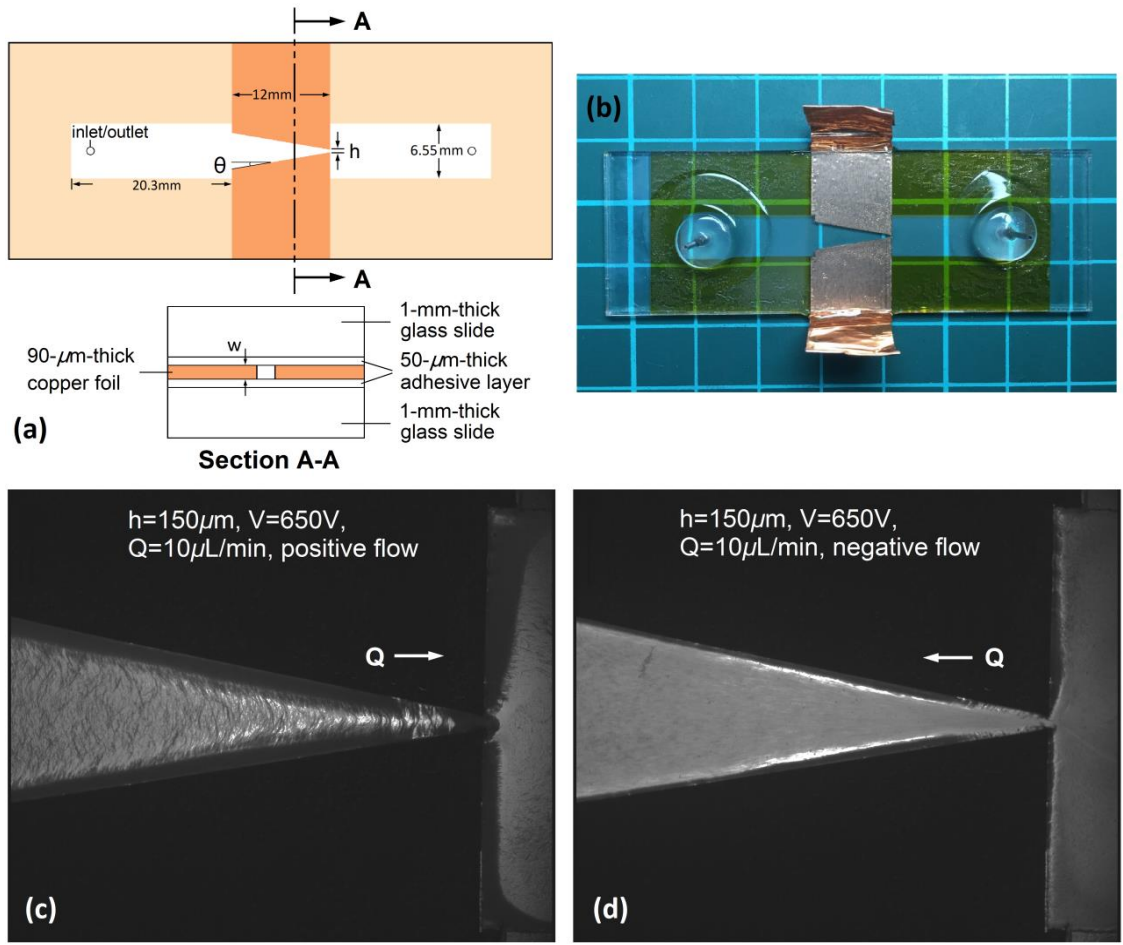

**FIG. 1.** Chip design and definition of the positive versus negative flow directions. (a) Scheme of the funnel-shaped electrode within the microfluidic channel, where the tilt angle of the electrode sides is defined as $\theta$, while the width of the funnel tip opening is defined as $h$. Chips of varying $h$ and $\theta$ were used while keeping the same average cross-section area of the funnel. (b) Image of the chip. (c) Microscopy image of the tip region under a flow of Q=10μL/min in the positive direction and an AC voltage of 650V and 10Hz square wave form. Particle blockage of the funnel tip is clearly seen. (d) Microscopy image of the tip region under the same conditions as in (c) with flow in the negative direction, where the hydrodynamic shearing of the particle chains leaves an open path for the ER fluid.

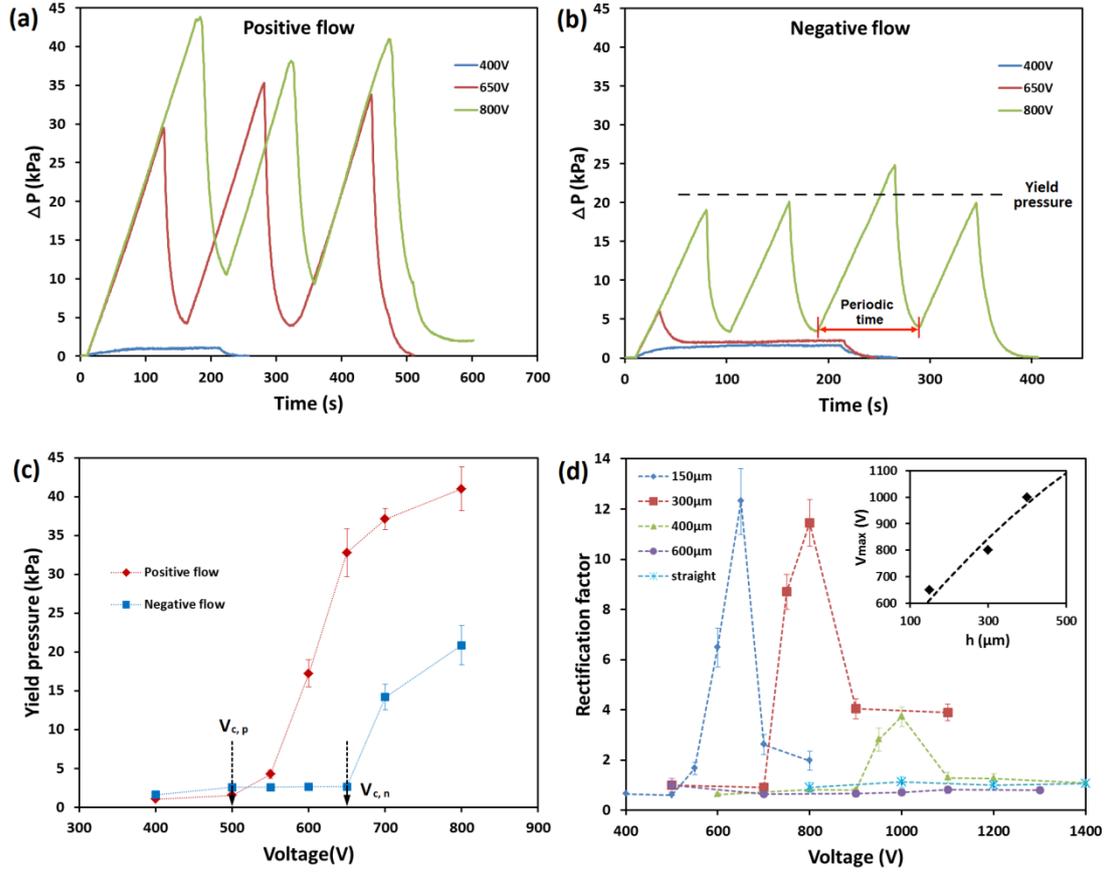

**FIG. 2.** The effect of the funnel geometry on ER rectification. Transient pressure drop response for (a) positive and (b) negative flow direction at different voltages (400V, 650V and 800V (10Hz frequency)). The width of the tip opening was $h=150\mu m$ and the flow rate was $Q=10$ $\mu L/min$. Beyond the critical voltage for particle blockage of the funnel tip, the response became oscillatory. (c) The yield pressure vs. voltage for both the positive and negative directions as obtained from parts (a) and (b), respectively. (d) The rectification factor vs. voltage for different widths (150, 300, 400, 600 and 2470$\mu m$ (straight channel)) of the funnel tip. With decreasing tip width, the voltage corresponding to the maximum rectification factor, $V_{max}$, decreased (see also inset; symbols represent experiments while dashed line represents theoretical scaling Eq.(3)), while the rectification factor increased.

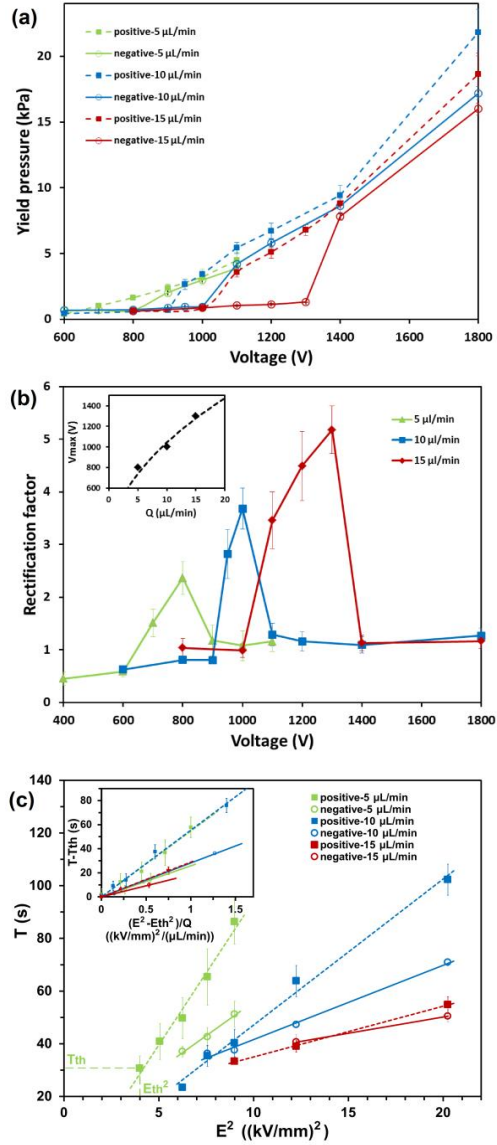

**FIG. 3.** The effect of the flow rate on ER rectification and pressure drop oscillations. (a) The yield pressures for both positive (dashed lines) and negative (continuous lines) flow directions vs. voltage, for flow rates of 5, 10 and 15$\mu$L/min. (b) The rectification factor vs. voltage for different flow rates of 5, 10 and 15$\mu$L/min. With increasing flow rates, the voltage corresponding to the maximum rectification factor, $V_{max}$, increased (see also inset; symbols represent experiments while dashed line represents theoretical scaling Eq.(3)), as well as the maximum rectification factor. (c) The time period of the pressure oscillations vs. the electric field squared for both positive and negative flow direction and varying flow rates. The inset depicts the universal scaling Eq. (4) of the measured data taken relative to the threshold value of the electric field ($E_{th}$) for inducing oscillations (each case has its own threshold value) and the corresponding threshold periodic time ($T_{th}$).

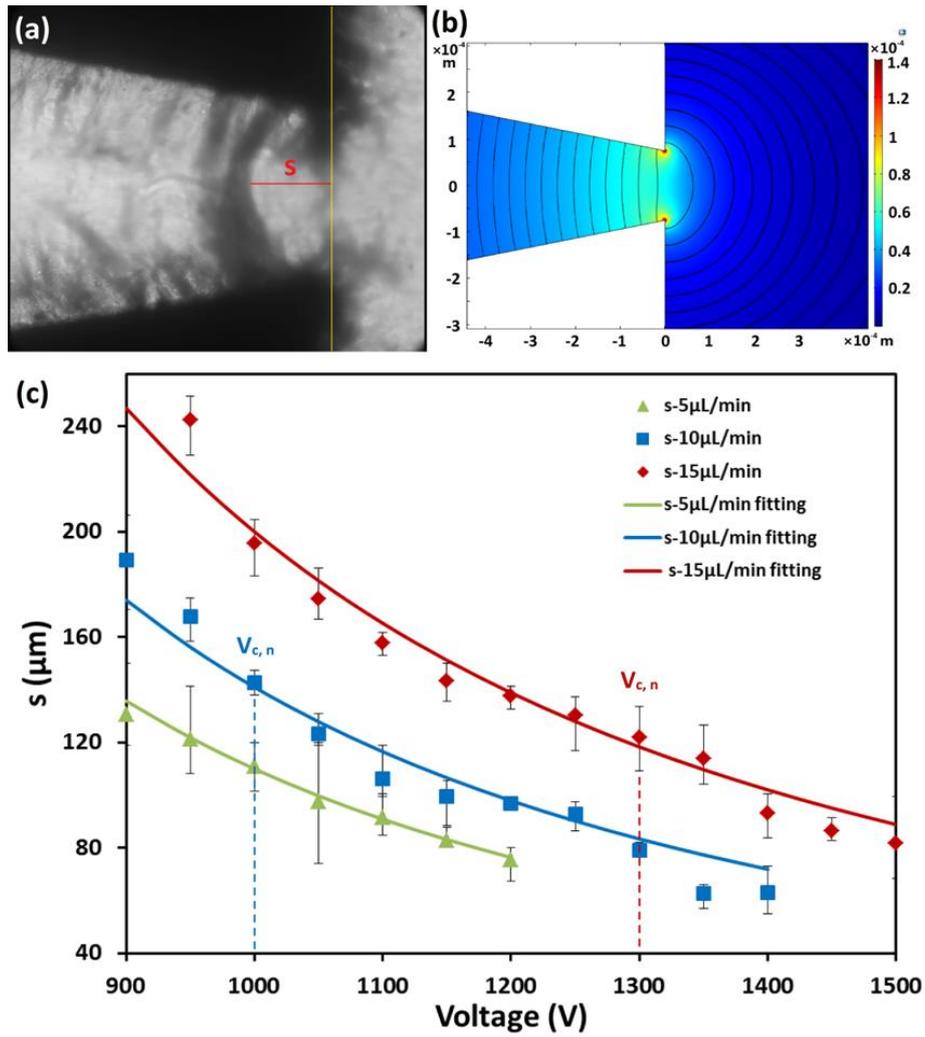

**FIG. 4.** The underlying physical mechanism of asymmetric ER response upon change of flow direction. (a) Microscopic image of a funnel with tip opening of 400μm defining the distance, *s*, where the chain is first formed under negative flow direction. (b) The electric field (norm), as calculated using numerical simulations indicating its intensification at the electrode corners. (c) The distance, *s*, is inversely proportional to $E^2$. The critical voltages for the negative flow direction, $V_{c,n}$, and their corresponding critical distance of chain formation, $s_c$. For $s>s_c$, the formed chains were able to escape the DEP trap, and thus, do not jam the entrance to the funnel.